\documentclass[twoside]{dis09}
\usepackage[latin1]{inputenc}
\usepackage[dvips]{graphicx,epsfig,color}
\usepackage{wrapfig,rotating}
\usepackage{amssymb,amsmath,array}
\usepackage{url}
\newcounter{comment}

\begin{document}
\title{DVCS and the skewness effect at small $x$
\footnotetext{ Talk given by D.M. at DIS 2009, 7--11 April 2008,
Madrid \cite{Mueller:DIS2009slides}.} }

\author{Kre\v{s}imir Kumeri\v{c}ki$^1$ and  Dieter M\"uller$^2$
%
%
\vspace{.3cm}\\
%
1- University of Zagreb - Department of Physics \\
P.O.B. 331, HR-10002 Zagreb - Croatia\\
\vspace{.1cm}\\
2- Ruhr-Universit\"at Bochum - Institut f\"ur Theoretische Physik II\\
D-44780 Bochum - Germany\\
}

\maketitle

\begin{abstract}
We analyze small-$x$ DVCS data using flexible GPD models and
compare our outcome with the full Shuvaev transformation. We point
out that the full Shuvaev transform  is a {\bf\em model} that
is equivalent to a {\em conformal} GPD  and a {\em minimalist}
``dual'' parameterization. Some mathematical subtleties of conformal
representations are recalled. We also comment on a speculation
of a factorization breakdown in DVCS.
\end{abstract}


Deeply virtual Compton scattering (DVCS), $\gamma^{*}(q_1)\,
p(P_1) \to \gamma(q_2)\, p(P_2)$,
is viewed as the cleanest process  to access generalized parton
distributions (GPDs), which encode a partonic description of the
nucleon structure. In the kinematics of H1 and ZEUS
collider experiments at HERA, the DVCS cross section is to a large
extent dominated by the flavor singlet part of the helicity conserved
Compton form factor (CFF), denoted as ${}^{\rm
S}\!\mathcal{H}$:
\begin{eqnarray}
\label{Def-CroSec1}
\frac{d\sigma}{d t}(W,t,{\cal Q}^2) \approx
\frac{4   \pi \alpha^2 }{{\cal Q}^4}\xi^2
\left| {}^{\rm S}\!\mathcal{H}\left(\xi,t=\Delta^2,{\cal Q}^2\right) \right|^2 \,.
\end{eqnarray}
Here, $W$ is the c.o.m.~energy, $\Delta=P_2 - P_1$ is the
momentum transfer, $-\mathcal{Q}^2=q_{1}^2$ is the
incident photon virtuality, and $\xi= {\cal Q}^2 / (2 W^2+{\cal
Q}^2)\approx x_{\rm Bj}/2$ is a Bjorken-like scaling variable.

The CFF ${}^{\rm S}\!\mathcal{H}$ factorizes further into a
convolution of the partonic, i.e., hard scattering, amplitude
$\mbox{\boldmath $C$}=({}^{\Sigma}C, {}^{G}\! C)$ and GPDs
$\mbox{\boldmath $H$}=({}^{\Sigma}H, {}^{G}\! H)$,
($\Sigma$=singlet quark, $G$=gluon),
\begin{equation}
{}^{\rm S}\!\mathcal{H}(\xi, t, \mathcal{Q}^2) =
\int_{-1}^1\! {\rm d}x\; \mbox{\boldmath $C$} (x, \xi, \mathcal{Q}^2/\mu^2,
\alpha_{s}(\mu)) \;
\mbox{\boldmath $H$} (x, \eta=\xi, t, \mu^2) \; ,
\label{eq:conv}
\end{equation}
where the skewness parameter $\eta=-\Delta\cdot q/(P_1+P_2)\cdot
q$ is set equal to $\xi$. The factorization scale $\mu$ separates
short- and long-distance dynamics and is often taken as
$\mu=\mathcal{Q}$. The scale dependence is governed by evolution
equations. Note that gluons do not directly enter the DVCS
amplitude at leading order (LO), but rather drive the evolution of
singlet quarks. Since the momentum fraction $x$ is integrated out
in (\ref{eq:conv}), GPDs cannot be directly revealed.

It is our objective to find flexible GPD models, which satisfy the
known theoretical constraints, and which can be pinned down by
fits to H1 and ZEUS DVCS data \cite{DVCSdat} at LO and beyond
\cite{KumMue09}. Thereby, we find it convenient to work with
conformal GPD moments. For {\em integral} conformal spin $j+2$
they are defined by  convolution with Gegenbauer polynomials
$C_j^{\nu} (x)$, e.g., for quarks:
\begin{equation}
H^{q}_{j}(\eta, t, \mu^2) \equiv \frac{\scriptstyle
\Gamma(3/2)\Gamma(j+1)}{\scriptstyle 2^{j+1} \Gamma(j+3/2)}\int_{-1}^1\!
{\rm d}x\; \eta^j\,
 C_j^{3/2} (x/\eta)\, H^{q}(x, \eta,  t, \mu^2) \;,
\label{eq:defconf}
\end{equation}
where $q$ is the flavor index. In the forward limit $\Delta\to 0$
these moments simply reduce for $j = 0,1,2,\cdots$ to familiar
$q_j(\mu^2)$ Mellin moments of parton distribution functions (PDFs). Conformal symmetry guarantees
that they evolve autonomously under evolution at LO (except for
the quark-gluon mixing). Now we require an appropriate behavior of
the conformal moments (\ref{eq:defconf}) for $j\to \infty$ with
$|{\rm arg}(j) |\le \pi/2$ \cite{MueSch05}. Carlson's theorem then states
that their analytic continuation with respect to $j$ is
unique and the GPD moments can be used to calculate the corresponding
CFF, cf.~(\ref{eq:conv}),
\begin{equation}
 \label{Def-MB-int}
 {^{\rm S}\!{\cal H}}(\xi,t,{\cal Q}^2)
 = \frac{1}{2i}\int_{c-i \infty}^{c+ i \infty}\!
 dj\,\xi^{-j-1} \left[i +
 \tan
 \left(\frac{\pi j}{2}\right) \right]
 \mbox{\boldmath $C$}_{j}({\cal Q}^2/\mu^2,\alpha_s(\mu))
 \mbox{\boldmath $H$}_{j}(\xi,t,\mu^2)
\,
\end{equation}
within a Mellin-Barnes integral
\cite{MueSch05,KumMuePas07,KumMue09}. Here the singularities of
the integrand, except for those of $\tan(\pi j/2)$ at
$j=1,3,\cdots$, lie on the l.h.s. of the integration path.

\begin{figure}[h]
\vspace{-3mm}
\centerline{
\includegraphics[scale=0.9,clip]{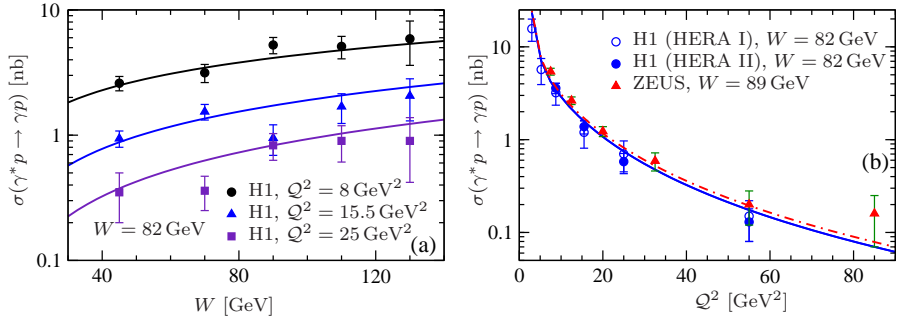}}
\vspace{-3mm}
\caption{\label{fig:DVCS} \small
LO fit to the DVCS cross section from  H1 and ZEUS
versus $W$ (left) and ${\cal Q}^2$ (right).
}
\end{figure}

To get flexible GPDs and CFF $\cal H$ and to make (loose) contact
with Regge terminology \cite{KumMuePas07,KumMue09}, we expand
their moments in terms of $t$-channel SO(3) partial waves (PWs)
$\hat{d}^{J}_{0,0}$ \cite{Pol98},
\begin{equation}
H^q_{j}(\eta, t, \mu_0^2) = \sum_{J=J^{\rm min}}^{j+1}
\frac{ h_{j}^J}{J - \alpha(t)}\,
\beta^q_J(t)\, \eta^{j+1-J} \hat{d}^{J}_{0,0}(\eta)
\;, \label{eq:so3}
\end{equation}
which are labelled by the $t$-channel angular momentum $J$. These
PWs can be expressed in terms of the familiar Legendre
polynomials, where the cosine of the scattering angle $\theta$ may
be approximated by $-1/\eta$ and which are normalized to one for
$\eta\to 0$. We include an effective ``pomeron'' pole at
$\alpha(t)=\alpha_0 + \alpha' t$ in the PW amplitudes and
parameterize the residual $t$-dependence $\beta_J(t)$ by a dipole
or exponential ansatz. The strengths $h^J_{j}$ of leading $J=j+1$
PWs and the intercept $\alpha\equiv\alpha(t=0)$ are obtained from
fits to some DIS data, and the remaining parameter space is
further reduced by either truncation of the sum (\ref{eq:so3}) or
its model dependent resummation, so-called $\Sigma$-PW model, and
is constrained by DVCS fits, which turn out to have good quality
$\chi^2/{\rm d.o.f.}\approx 1$, see Fig.~\ref{fig:DVCS}. (We take
$\mu_0 = 2\, {\rm GeV}$ as input scale.)
One important partonic quantity that can be extracted from such
fits is the skewness ratio of GPDs, which is given by the value of
the skewness function at $\vartheta=\eta/x=1$:
\begin{eqnarray}
r(x,\vartheta,\mu^2) \equiv
\frac{H(x,\eta=\vartheta x,t=0,\mu^2)}
{H(x,\eta=0,t=0,\mu^2)}\,.
\label{eq:SkeRat}
\end{eqnarray}

Before we present results of our fits, some issues should be
clarified. Let us recall that there are at least four different
conformal GPD representations that are general and are used
in phenomenology: Shuvaev-Noritzsch%
\footnote{This integral transformation was proposed by
A.~G.~Shuvaev in \cite{Shu99}. The term {\em (full) Shuvaev
transformation} was used in
Refs.~\cite{ShuBieMarRys99,MarNocRysShuTeu08} for an integral
transformation that results in a \emph{restricted} GPD transform.
This restriction was removed by J.~Noritzsch in Ref.~\cite{Nor00}.
He was also the first who utilized this integral transformation to
set up flexible GPD models, describing small-$x$ DVCS data at LO.
Thus, --- not only to avoid confusion --- we name the
\emph{general} integral transformation Shuvaev-Noritzsch.}
transformation \cite{Shu99,Nor00}, ``dual" parameterization
\cite{PolShu02}, and two versions of Mellin-Barnes representations
\cite{MueSch05,KirManSch05}. All of them start from an integral
conformal SL(2 $|\mathbb{R}$) PW expansion of the (crossed) GPD in
terms of Gegenbauer polynomials (now viewed as generalized functions in the mathematical sense),
\begin{eqnarray}
H^q(x,\eta,t) =\! \sum_{j=0}^\infty  \eta^{-j-1} p_j(x/\eta) H^q_j(\eta,t)\,,
\;\;  p_j(x) =\theta(1\!-\!|x|)\frac{2^j \Gamma(5/2+j) (1-x^2) }{\Gamma(3/2)\Gamma(3+j)} C^{3/2}_j(x)\, ;
\end{eqnarray}
however, in the end they provide
different representations for GPDs. Since these specific conformal
representations involve some mathematical subtleties, efforts to
understand them have been undertaken in
Refs.~\cite{Nor00,MueSch05,KirManSch05,Pol07,KumMue09}. We
consider the one-to-one correspondence among the various
representations established already by construction, as sketched
\begin{wrapfigure}{r}{0.45\columnwidth}
\vspace{-3mm}
\centerline{
\includegraphics[scale=0.55,clip]{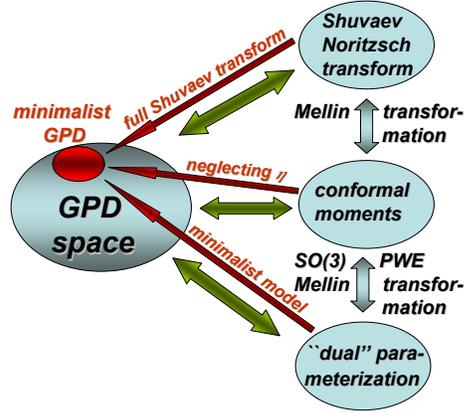}
}
\vspace{-3mm}
\caption{\label{fig:Transformations} \small
The one-to-one relation of various conformal GPD
representations is indicated by thick arrows.
The restriction of minimalist GPD models is sketched by thin
arrows.}
\end{wrapfigure}
by the thick arrows in Fig.~\ref{fig:Transformations},
although the inverse transformations are only partly known.
Furthermore, if some mathematical subtleties are ignored, general
conformal representations {\em degenerate} to {\em parameter-free}
GPD models for the small-$x$ region.

In the small-$x$ region the one-to-one correspondence among the
different parameter-free models can be analytically shown by
noting that they all have the conformal value of skewness
function, e.g., for $t=0$ and fixed ${\cal \mu}^2$
\begin{eqnarray}
r^{\rm con}(\vartheta,\mu^2)
\simeq
{_2F_1}\!\left(\!{\alpha/2,(1+\alpha)/2 \atop
3/2+\alpha}\Big|\vartheta^2\!\right).
\label{eq:ConSkeRat}
\end{eqnarray}
Assuming, like in most standard PDF parameterizations, that {\em
effective} Regge behavior is included in the intercept
$\alpha\equiv\alpha(t=0)$, one easily finds the skewness function
(\ref{eq:ConSkeRat}) from the Mellin-Barnes representation
\cite{MueSch05,KirManSch05} of a conformal GPD model
\cite{KumMuePas07} by shifting the original integration path to
the l.h.s. in the complex $j$-plane, cf.~Eq.~(\ref{Def-MB-int}).
Thereby, the hypergeometric function in (\ref{eq:ConSkeRat}) is
nothing but the Clebsch-Gordan coefficient of the conformal PW
expansion, taken at $j=\alpha-1$. As long as the conformal moments
behave smoothly in the vicinity of $\eta=0$, a GPD model
degenerates to a conformal one. For the full Shuvaev
transformation model the skewness function (\ref{eq:ConSkeRat})
can be found in Ref.~\cite{MarNocRysShuTeu08} as the standard
integral representation of hypergeometric functions,
\begin{eqnarray}
r^{\rm con}(\vartheta,\mu^2)
\simeq
\frac{2^{2\alpha+1} \Gamma(3/2+\alpha)}{\Gamma(1/2)\Gamma(1+\alpha)}\,
x^{\alpha}  \int_{0}^1\! ds\,
s^\alpha (1-s)^\alpha (x+\xi-2 \xi\, s)^{-\alpha} \Big|_{\xi = \vartheta x}
\,.
\end{eqnarray}

For an effective ``pomeron'' ($\alpha \sim 1$) and ``Reggeon''
($\alpha \sim 1/2$) pole such analytic approximation works quite
well for $x=\xi \lesssim 10^{-2}$. Further singularities  ($0<
\alpha \lesssim 1$) might be taken into account, too. If the
scale grows the effective ``pomeron'' intercept $\alpha$
increases, according to the well-known double-log approximation.
If ``Reggeon'' contributions still play a role, the analytic
approximation of a GPD will of course be a linear combination of
conformal skewness functions (\ref{eq:ConSkeRat}).

As the reader realizes, for the same small-$x$ GPD model two
different names are used, or even three if one includes the {\em
minimalist} ``dual'' parameterization. In the remaining part of
this presentation we simply adopt the terminology from the
``dual'' parameterization \cite{PolShu02} and use, instead of the
term ``parameter-free  prediction'', the term ``minimalist GPD'' for
both our conformal and the full Shuvaev transformation model.
Let us now scrutinize some statements presented in
Ref.~\cite{MarNocRysShuTeu08}, and used there to support the
problematic idea that the minimalist GPD should be considered as
more than just a restricted GPD model \cite{DieKug07a}.
\vspace{1mm}

\noindent {\em i.} A naive Taylor expansion of integral GPD
moments around $\eta=0$  would imply that the Shuvaev-Noritzsch
GPD transform reduces to a PDF at small $\eta$. However, it is
well known that the operations of the small-$\eta$ expansion and
of an analytic continuation of polynomials with respect to their
order do not commute. The Sommerfeld-Watson transformation of a
series, representing a GPD in terms of integral moments, is only
possible if the assumptions of the Carlson theorem are
satisfied. The theorem then states that the continuation of
conformal moments is unique. If the integral GPD moments are first
reduced to PDF ones, the Sommerfeld-Watson transformation  is not
applicable \cite{MueSch05}. That the Shuvaev-Noritzsch GPD
transform might not reduce to a PDF is known since
Refs.~\cite{MusRad00,Nor00}. The analogue of this in the ``dual''
parameterization framework is the increase of singular behavior of
non-leading forward-like functions \cite{Pol07}.

\begin{wrapfigure}{r}{0.5\columnwidth}
\vspace{-8mm}
\centerline{
\includegraphics[scale=0.6,clip]{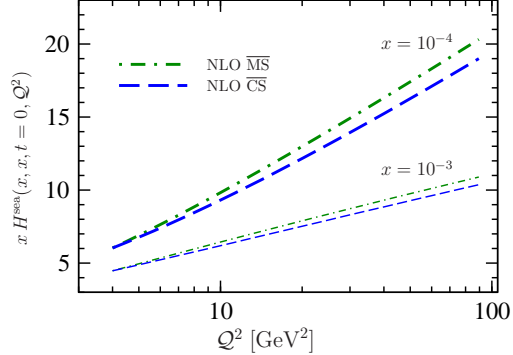}}
\vspace{-3mm}
\caption{\label{fig:CSvsMS} \small GPD on the cross-over line versus
${\cal Q}^2$ in the $\overline {\rm MS}$ scheme
(dot-dashed) and the $\overline {\rm CS}$ one (dashed).
The discrepancy is the error
if a full Shuvaev transform is used in the $\overline{\rm MS}$ scheme.
}
\end{wrapfigure}
\noindent
{\em ii.} It was said in \cite{MarNocRysShuTeu08} that the absence of
poles in the right half of complex
$j$ plane is a physical requirement that is sufficient to guarantee that
the full Shuvaev tranformation (minimalist) model
is the unique GPD model at small $\eta$ to $O(\eta)$ accuracy.
We first remark that such a requirement would be
\emph{physical} only for the angular momentum $J$. Moreover,
counterexamples are known, such as resummed
$\Sigma$-PW model used here and in \cite{KumMue09} for DVCS fits.
A small-$\eta$ expansion reveals then that the poles in the right half of complex
$j$ plane are an artifact of the expansion itself and, thus, should be considered as
spurious \cite{KumMue09}.
\vspace{1mm}

\noindent
{\em iii.} It was stated in \cite{MarNocRysShuTeu08} that up to
order $O(\eta)$ the full Shuvaev transformation  is compatible with the
NLO evolution equation in the $\overline{\rm MS}$ scheme. However,
this is not true, as spelled out, e.g., in \cite{KumMuePas07},
and demonstrated in Fig.~\ref{fig:CSvsMS}, where we show the NLO
evolution of a GPD on $\eta=x$ both in the
$\overline{\rm MS}$ scheme and within the procedure \cite{MarNocRysShuTeu08},
which actually corresponds to an evolution operator in
$\overline{\rm CS}$ scheme,
as dot-dashed and dashed curves, respectively.
One realizes that the discrepancy increases with evolution and
with decreasing $x$. Hence, it cannot be an $O(\eta)$ effect.

\vspace{1mm}

\begin{figure}[t]
\vspace{-24mm} \centerline{
\includegraphics[scale=0.6,clip]{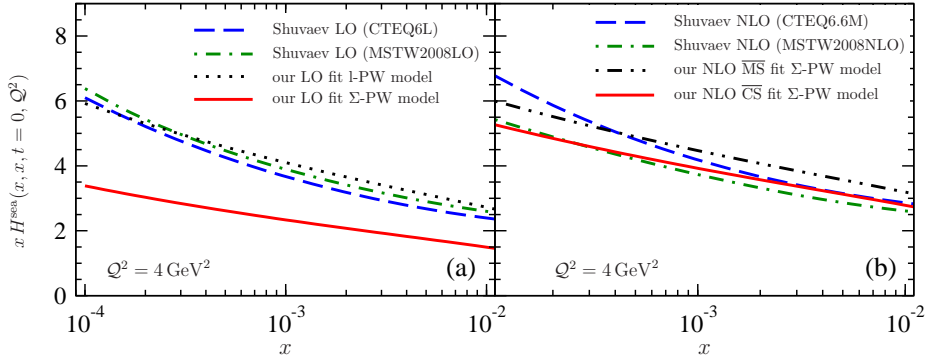}}
\vspace{-4mm} \caption{\label{fig:Shuvaev} \small Sea quark GPDs
on the cross-over line obtained from good fits in
$\overline {\rm CS}$ (solid) and $\overline {\rm MS}$ (dot-dot-dashed) schemes
are compared with minimalist GPD models used in our bad
${\chi^2}/{\rm d.o.f.}\approx 3$ LO fit (dotted) and from Ref.~\cite{MarNocRysShuTeu08}
(dashed, dot-dashed)  at LO (left) and NLO (right).
}
\vspace{-3mm}
\end{figure}

The phenomenological status of a minimalist GPD model was to a
large extent discussed in Ref.~\cite{KumMue09}. It is shown in
Fig.~\ref{fig:Shuvaev}a that a flexible model (solid), pinned down
by a good  $\chi^2/{\rm d.o.f.}\approx 1$ fit, is almost two times
smaller than a minimalist GPD model, shown by the other curves,
and so the latter model is ruled out at LO. This was also
confirmed within the ``dual'' parameterization and standard PDFs
\cite{GuzTec08}. (Small shape discrepancy of our minimalist model
(dotted) and those of Ref.~\cite{MarNocRysShuTeu08} (dot-dashed,
dashed) comes from our neglecting of ``Reggeon'' contributions.)
At NLO one might conclude from Fig.~\ref{fig:Shuvaev}b that within
experimental and theoretical uncertainties the minimalist models
\cite{MarNocRysShuTeu08} (dot-dashed, dashed) are compatible with
ours (solid, dot-dot-dashed), obtained from data. Still, a more
detailed view shows that skewness effect, $t$-dependence, and
scheme convention are interrelated. For instance, we found that in
$\overline{\rm MS}$ scheme the $\Sigma$-PW model with dipole
$t$-dependence (dot-dot-dashed) is somewhat away from a minimalist
one (which provided only $\chi^2/{\rm d.o.f.}\approx 1.5$). On the
other hand in the $\overline{\rm CS}$ scheme at NLO (solid) and
NNLO it turned out to be close to a minimalist one. We add that
the error from the wrong evolution, see Fig.~\ref{fig:CSvsMS}, is
comparable to the PDF uncertainties and that for the kinematical region
of interest even the forward evolution operator is not stable in
perturbation theory. Note that the reparameterization effects for
our quark GPD models, resulting from switching on NLO corrections,
are much larger than for PDFs. We view this as a sign that in our
modelling we have not yet reached control over evolution.

In conclusion, the full Shuvaev transform is a minimalist  GPD
model valid at small-$x$, which is also realized in other versions
of conformal representations. So far we see no general theoretical
arguments that support preference of this model over others. NLO
evolution is performed in Ref.~\cite{MarNocRysShuTeu08} in such a
way that it is inconsistent with the $\overline{\rm MS}$ scheme. A
minimalist model might describe DVCS data at NLO, however, not
necessarily when precision level is reached. Finally, we recall
that the skewness effect for gluons is in all popular GPD models
approximately the same, namely, zero (i.e. $r^{\rm
G}(\vartheta=1)\approx 1$; what is often quoted as a large
skewness effect is 
$2^{\alpha-1}$) \cite{KumMue09}. Hence, the skewness model
uncertainty, compared to  the phenomenological PDF uncertainty,
might presumably be considered as small at NLO. Fully flexible GPD
models, allowing the control over evolution, should be invented
and confronted with DVCS data in further studies. We hope that
more DVCS HERA II run data will become available and so the
statistical errors will be reduced. This might then allow to
address the observables and partonic quantities, as discussed in
Ref.~\cite{KumMue09}.

\section*{Acknowledgments}
This work was supported in part by the BMBF (Federal Ministry for
Education and Research), contract FKZ 06 B0 103 and by the
Croatian Ministry of Science, Education and Sport under the
contracts no. 119-0982930-1016 and 098-0982930-2864. For some
mathematical insights we are indebted to J.~Noritzsch, we are grateful
to M.~Diehl for his effort to clarify the issues, and we thank
A.~D.~Martin, M.~G.~Ryskin, and T.~Teubner for discussions.

\section*{Comment added}
At the DIS 2009 conference Felipe J.~Llanes-Estrada presented, as
collaborator of S.J.~Brodsky, J.~Tim Londergan, and Adam P.
Szczepaniak, a talk \cite{Estrada:DIS2009slides} with the title
{\em ``Regge behavior in DVCS: non-factorizability and J = 0 fixed
pole''} in which it was stated that Regge behavior is responsible
for the breakdown of collinear factorization. D.M.~likes to take
the opportunity here to repeat his comment, given after the DIS
2009 summary talk {\em ``Structure Functions and Low-$x$''}, that
this speculation is based on a ``bad'' GPD model. Further
clarifications, e.g., that the illness of this ``bad'' GPD model
arises from the ultraviolet/infrared region rather than from the
Regge behavior, how to formulate a ``good'' GPD model in the
reggeized parton model framework, and how to fix a ``bad'' GPD
model within analytic regularization, are presented in
Ref.~\cite{KumMuePas07a}. Interestingly, the authors find it not
legitimate to use analytic regularization to fix their model,
however,  suggest the same regularization to extract a so-called
$J=0$ pole within the GPD formalism that, according to them,
cannot be applied to DVCS (i.e., small $-t$). In contrast to the
collinear factorization approach \cite{KumMue09}, it could not be
demonstrated so far that the Regge exchange amplitude model,
introduced two years ago, describes DVCS data. In our opinion the
reason is simple, namely, the Regge exchange amplitude model is
already ruled out by H1 and ZEUS data \cite{KumMuePas07a} and
maybe also by the fact that the scaling hypothesis works rather
well in a global fit to fixed target DVCS measurements
\cite{KumMue09}. Thereby, a `partonic dispersion relation'
approach is utilized. Since some clarification is still needed,
see conclusion in Ref.~\cite{GolLui09}, we like to spell out
again, see Refs.~\cite{Ter05,KumMuePas07,KumMuePas07a} and
references therein, that our uses of `partonic dispersion
relations' is  in {\em one-to-one} correspondence to the collinear
factorization approach, taken in the LO approximation and
supplemented with the scaling hypothesis.



\begin{footnotesize}

\end{footnotesize}


\end{document}